\documentclass{iopart}

\usepackage{iopams}
\usepackage{graphicx}

\newcommand{\ket}[1]{\vert #1\rangle}

\newcommand{\binom}[2]{ {#1 \choose #2} }
\newcommand{\Jx}{J_{ex}}

\newcommand{\Hh}{\mathcal{H}_D}

\begin{document}

\title[Unexpected systematic degeneracy in coupled Gaudin models]{Unexpected systematic degeneracy in a system of two coupled Gaudin models with homogeneous couplings}

\author{B.\, Erbe and J.\, Schliemann}

\address{Department of Physics, University of Regensburg}

\begin{abstract}
We report an unexpected systematic degeneracy between different multiplets 
in an inversion symmetric system of two coupled Gaudin models with 
homogeneous couplings, as occurring for example in the 
context of solid state quantum information processing. We construct the 
full degenerate subspace (being of macroscopic dimension), which turns out to lie in the kernel of the 
commutator between the two Gaudin models and the coupling term. 
Finally we investigate to what extend the degeneracy is related to the 
inversion symmetry of the system and find that indeed there is a large
class of systems showing the same type of degeneracy. 
\end{abstract}


\maketitle
\section{Introduction}

In a large variety of nanostructures spins couple to a bath of other 
spin degrees of freedom. Commonly such systems are  described by so-called 
central spin models. 
Important examples are given by semiconductor
\cite{Petta06, Koppens06, Hanson07, Braun05} and carbon nanotube 
\cite{Churchill09} quantum dots, phosphorus donors in silicon \cite{Abe04}, 
nitrogen vacancy centers in diamond \cite{Jelezko04, Childress06, Hanson08} 
and molecular magnets \cite{Ardavan07}. Motivated by the perspective to 
utilize the central spins as qubits \cite{Loss98, Kane98} or to effectively 
access the bath spins via the central spins \cite{Schwager1, Schwager2}, 
presently central spin models are the subject of extensive theoretical as 
well as experimental research. However, their importance in a more 
mathematical context became clear already several decades ago, when 
Gaudin proved the Bethe ansatz integrability of the central spin model 
with one central spin (Gaudin model) \cite{Gaudin76}. Since then they are 
in the focus of the important field of quantum integrability \cite{Garajeu, Sklyanin96, Frenkel04, ErbS09, John09, ErbSl10}.

It is well-known that the energy levels of a quantum system usually tend to 
repel each other and degeneracies are exceptional events \cite{Neumann29}. 
Hence there are only extremely few examples of systems with degenerate 
eigenstates and \textit{even less}, whose eigenstates are 
\textit{systematically} degenerate. Famous examples are given by the 
hydrogen atom \cite{Pauli26}, the $n$-dimensional harmonic oscillator 
\cite{Baker56} or the Haldane-Shastry model \cite{Haldane92, Frahm93}. 
In all three cases the degeneracies are due to hidden symmetries requiring a
dedicated analysis.

In the present work we  we study the spectrum of an inversion symmetric 
central spin model consisting of two coupled Gaudin models with homogeneous 
coupling constants, meaning they are chosen to be equal to each other.
In order to lower the dimension of the problem, the baths of the two Gaudin 
models are approximated by single long spins, which does not change the 
set of eigenvalues of the Hamiltonian.
Surprisingly, the resulting Hamiltonian exhibits systematically degenerate 
multiplets of consecutive
total angular momentum and alternating parity, a situation somewhat similar
to the degenerate multiplets of orbital angular momentum in the hydrogen
atom. 

The outline of the paper is a follows.
The degeneracies in the coupled Gaudin models are first analyzed in a numerical
approach in Sec.~\ref{model}. In Sec.~\ref{constr} we analytically
construct the full subspace of degenerate states which turns out to be 
located in the kernel of the commutator between the two Gaudin models and the 
coupling term. In Sec.~\ref{inv} we furthermore investigate the role of the 
inversion symmetry and show that indeed there is a whole class of 
systems with spectra showing the same type of degeneracy.

\section{Model and spectral properties}
\label{model}

The Gaudin model \cite{Gaudin76} describes the coupling of a central spin 
$\vec{S}_i$ to a set of $n_i$ bath spins $\vec{I}_i^j$
\begin{equation}
\label{Gaudin}
 H_i=\vec{S}_i \sum_{j=1}^{n_i} A_i^j \vec{I}_i^j,
\end{equation}
via some coupling constants $A_i^j$, which have the unit of energy. 
In the following we choose the couplings to be homogeneous, 
i.e. $A_i^j=A_i$. In this case the central spin couples to a simple sum of 
spins, denoted by $\vec{I}_i$ from now on. Furthermore we assume 
$S_i=1/2$. Coupling together two such Gaudin models $H_G:=H_1+H_2$ by
\begin{equation}
 H_c=\Jx \vec{S}_1 \vec{S}_2
\end{equation}
finally yields the Hamiltonian subject to our investigation:
\begin{equation}
\label{H}
\nonumber H=H_G+H_c= A_1 \vec{S}_1 \vec{I}_1+ A_2 \vec{S}_2 \vec{I}_2 + \Jx \vec{S}_1 \vec{S}_2.
\end{equation}
The parameter $ \Jx$ can be viewed as an exchange coupling.
Obviously the Hamiltonian conserves the total spin 
$\vec{J}=\vec{S}_1 + \vec{S}_2 + \vec{I}_1 + \vec{I}_2$ as well as 
$\vec{I}^2_1$ and $\vec{I}^2_2$.

The $n_i$ bath spins couple to different values $I_i$ of the total bath 
spin squared. In the following we study the spectrum of 
the Hamiltonian for $A_1=A_2=A$, where $A:=(1/2)(A_1+A_2)$, on subspaces 
$I_1=I_2=:I$. On these subspaces, 
in addition to the symmetries mentioned above, $H$ is invariant under 
``inversions'', meaning an interchange $1 \leftrightarrow 2$. 
It is clear that this is not the case globally, i.e. on the entire Hilbert 
space. However, subspaces 
with $I_1=I_2$ lie fully in the kernel of the commutator 
$\left[H,\tau\right]$, where $\tau$ denotes the inversion operator. 
Obviously this only has the two eigenvalues $(\pm 1)$. In the following we 
refer to this as positive and negative parity. 

In order to reduce the dimension of the problem, we approximate each 
bath by one single spin of length $I$. This neglects the quantum numbers 
associated with a certain Clebsch-Gordan decomposition of the respective 
bath and therefore the multiplicity of the eigenvalues changes, but not the set
of eigenvalues itself. 
Every energy in the resulting spectrum indeed appears $x_1x_2$ times in the 
spectrum of $H$, where $x_i$ denotes the number of multiplets 
with the quantum number $I_i$. If for example $I_i^j=\frac{1}{2}$, 
we have \cite{SKhaLoss03}
\begin{equation}
x_i = \left[  \binom{n_i}{\frac{n_i}{2}-I_i}
-\binom{n_i}{\frac{n_i}{2}-I_i-1}\right].
\end{equation}
However, it should be stressed again that the energy eigenvalues themselves remain unaltered. 

\begin{figure}
\includegraphics[width=8cm]{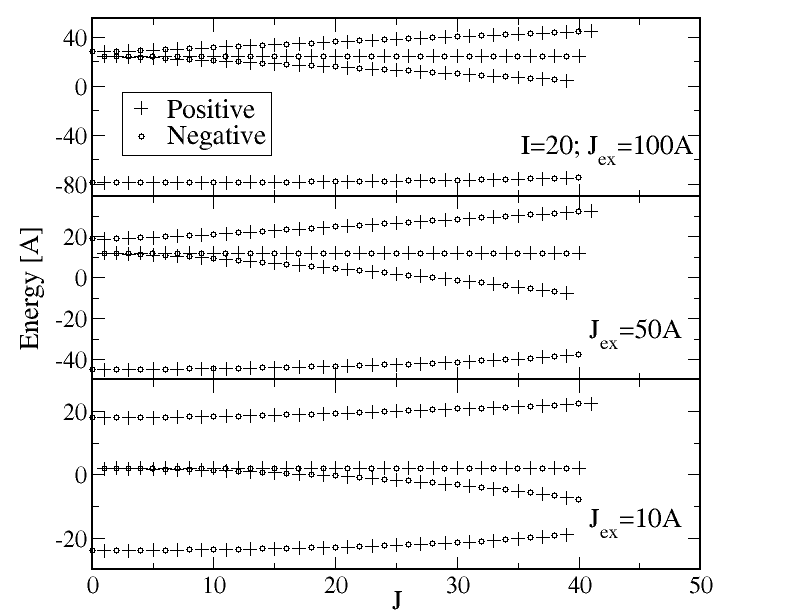}
\caption{\label{Fig:Spektrum1} Spectrum of $H$ with baths approximated 
by two single spins of equal length $I=20$ for different values of the 
exchange coupling $\Jx$. The energies (in units of $A$)
are plotted against the total spin $J$, i.e. each data point represents a 
multiplet of $2J+1$ states. States of positive (negative) parity are signalled
by a cross (circle).
For all exchange couplings $\Jx$ we have four ``branches'' of energy levels, 
where the above three ones originate from triplet states with respect to 
$H_c$ and the lower one is associated with the singlet state. 
The triplet branch of intermediate energy consists completely degenerate
multiplets of alternating parity.}
\end{figure}

\begin{figure}
\includegraphics[width=8cm]{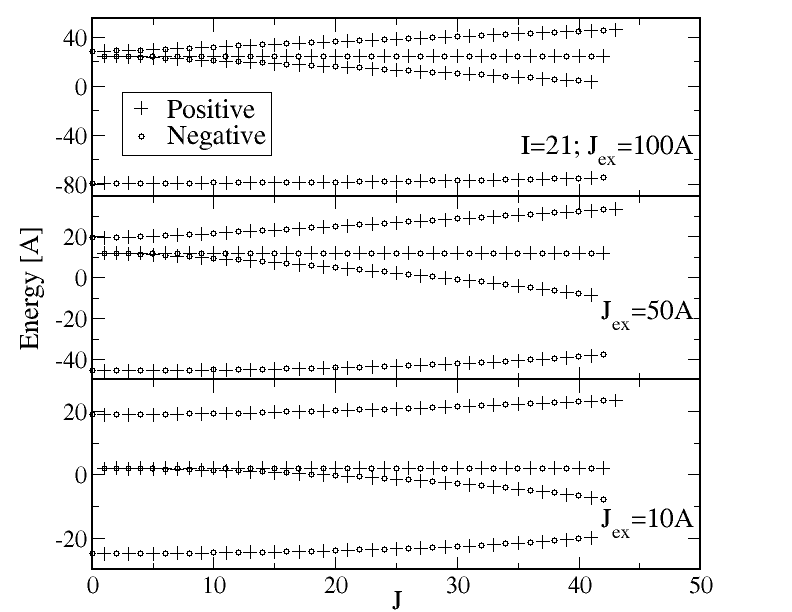}
\caption{\label{Fig:Spektrum2} Analogous data as in Fig.~\ref{Fig:Spektrum1} 
for an odd spin length $I=21$. Again we find a completely degenerate triplet
branch.}
\end{figure}

In Figs. \ref{Fig:Spektrum1} and \ref{Fig:Spektrum2} we show spectra
obtained numerically for different values of the 
exchange coupling constant $\Jx$, both for an even and an odd $I$. 
Although the spectra are quite rich in detail, their global structure
becomes already plausible from simple qualitative arguments.
Obviously we always 
have four ``branches'' of energy levels, where, in particular for large
$\Jx$, three of them form a bundle separated from the fourth one.
The three former branches consist of states where the two central spins
are predominantly coupled to a triplet (which has eigenvalue $\Jx /4$
under $H_c$) while in the latter branch the central spins are mainly
in the singlet state (having eigenvalue $-3\Jx /4$
under $H_c$).
The coupling of the central spin triplet and singlet to the bath spins
leads then leads to the observed further energy splittings between and within
the corresponding branches.

An unexpected particular feature, however, occurs in the triplet branch
of intermediate energy. Here all multiplets are energetically
completely degenerate with eigenvalue $(\Jx-2A)/4$. These multiplets have
consecutive total spin between $J=1$ and $J=2I$ and alternating
parity. Here positive (negative) parity corresponds to $2I-J$ being even (odd).
The latter observation is reminiscent to the degenerate multiplets of orbital 
angular momentum found in the hydrogen problem.
In general, such systematic degeneracies are extremely rare, and hence our 
finding is interesting on its own right. 
Moreover, since the degenerate subspace is of particularly high dimension, 
potential applications in, for example, solid state quantum 
information processing can be envisaged: 
It is clear that states with overlap exclusively in a 
degenerate subspace do not show any non-trivial time evolution. 
Therefore, such spaces have the potential to provide 
valuable implementations of long lived quantum memory, 
where the present one appears to be particularly 
suitable due to its enormous size. Note that even in the ``thermodynamic limit'' $I \gg 1$
a fourth of the Hilbert space is degenerate: The dimension of the full Hilbertspace is
$4(2I+1)^2$ and the degenerate subspace $\Hh$ has dimension $\sum_{n=1}^{2I}(2n+1)=4I(I+1)$,
yielding
\begin{equation*}
\frac{I(I+1)}{(2I+1)^2} \approx \frac{1}{4},
\end{equation*}
if $I \gg 1$. Furthermore, 
the space of degenerate states detected here decomposes into subspaces 
of different parity which could also serve as a computational basis for quantum 
information processing. 

\section{Construction of the degenerate subspace}
\label{constr}

So far we have reported on numerical observations revealing an unexpected
systematic degeneracy in the spectrum. In the following we analytically construct the subspace $\Hh$ of 
these degenerate multiplets.

\subsection{General ansatz and first consequences}

As we shall see below, the degenerate states are
simultaneous eigenstates of the Gaudin part $H_G$ of the Hamiltonian and
the coupling $H_c$ between the central spins. In other words, 
$\Hh$ lies entirely in the kernel of the commutator 
\begin{equation}
\label{Comm}
 \left[H_G,H_c \right] 
= -iA \Jx \left( \vec{S}_1 \times \vec{S}_2 \right) \cdot \left(\vec{I}_1 
- \vec{I}_2 \right)\,.
\end{equation}

Let us first turn to a single Gaudin Hamiltonian,
$H_i=A\vec S_i\vec I_i$, with $S_i=1/2$ and $I_i=I$.  The eigenvalues read
\begin{equation}
E_{\pm}(A,I)=\frac{A}{2}\left(\pm\left(I+\frac{1}{2}\right)+\frac{1}{2}\right)\,,
\end{equation}
and the eigenstates are given by a well-known
Clebsch-Gordan decomposition \cite{Schwabl}
\begin{equation}
\label{zstates}
\nonumber \ket{I \pm \frac{1}{2},m_i}=\mu^{\pm}(m_i)\ket{\uparrow}
\ket{I,m_i-\frac{1}{2}} \pm \mu^{\mp}(m_i)\ket{\downarrow}\ket{I,m_i+\frac{1}{2}},
\end{equation}
where, apart from standard notation, we have introduced
\begin{equation}\label{Mu}
 \mu^{\pm}(m)= \sqrt{\frac{I \pm m + \frac{1}{2}}{2I+1}}\,.
\end{equation}
The eigenvalues of $H_G=H_1+H_2$ now follow immediately 
\numparts
\begin{eqnarray}
 H_G \ket{ + ,m_1}\ket{ + ,m_2}&=& 
\frac{A}{2}(I_1+I_2)\ket{ + ,m_1}\ket{ + ,m_2} \\
 H_G \ket{ + ,m_1}\ket{ - ,m_2}&=& 
-\frac{A}{2}\ket{ + ,m_1}\ket{ - ,m_2}\\
 H_G \ket{ - ,m_1}\ket{ + ,m_2}&=& 
-\frac{A}{2}\ket{ - ,m_1}\ket{ + ,m_2}\\
 H_G \ket{ - ,m_1}\ket{ - ,m_2}&=& 
-\frac{A}{2}(I_1+I_2+2)\ket{ - ,m_1}\ket{ - ,m_2},
\end{eqnarray}
\endnumparts
where we abbreviated $\ket{I\pm \frac{1}{2},m_i}=:\ket{\pm,m_i}$. 
Obviously, the states $\ket{\pm,m_1}\ket{\mp,m_2}$ 
are degenerate with the eigenvalue being independent of $I$. 
As seen above, the highly degenerate eigenvalue in the subspace 
$\Hh$ is $(\Jx -2A)/4$. Thus, eigenstates of $H$ with this eigenvalue can be 
constructed by simply combining the states 
$\ket{\pm,m_1}\ket{\mp,m_2}$ to triplet states with respect to the two 
central spins, 
meaning that they lie in the kernel of (\ref{Comm}). At this point it is of 
course not clear that all eigenstates with the above eigenvalue are resulting 
through this approach. However, we will see that this is indeed the case.
\begin{figure}
\includegraphics[width=8cm]{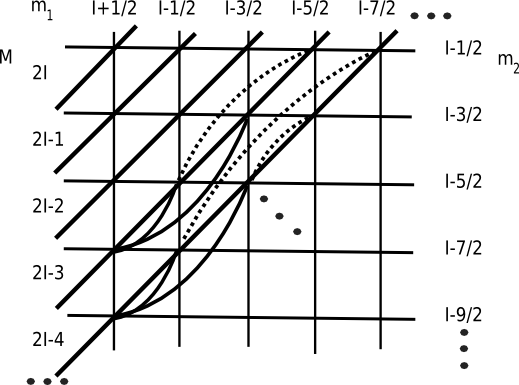}
\caption{\label{Fig:State} The different values of $m_1$ and $m_2$ 
arranged on a grid. The diagonal lines mark the states with constant 
magnetization. The dotted bended lines connect the states with 
interchanged magnetization, combined in our ansatz (\ref{ansatz}). 
The solid bended lines connect the states which are combined in order 
to construct the set of linearly independent eigenstates spanning the 
eigenspace to the eigenvalue $(J-2A)/4$.}
\end{figure}

In other words, our goal is to eliminate singlet contributions from 
suitable linear combinations of the states $\ket{\pm,m_1}\ket{\mp,m_2}$.
To this end we use an ansatz already accounting for 
the conservation of $J^z$ and the parity symmetry by superimposing states of
the form
\begin{equation}
\label{symmst}
\ket{+,m}\ket{-,M-m}\pm \ket{-,M-m}\ket{+,m}\,,
\end{equation}
where $M$ is the eigenvalue of $J^z$. All considerations will focus on $M \geq 0$, as
states with $M<0$ result simply by reversing every spin.
In the following analysis one needs to distinguish the four different 
cases depending on whether $M$ is even or odd, and $I$ is integer or
half-integer. This case-by-case procedure
can be nicely encapsulated and simplified as follows by 
introducing $i=2I-M$ with $i=0,\ldots, 2I$: 
In Fig.~\ref{Fig:State} the possible values of $m_1$ and $m_2$ are arranged on 
a grid. The diagonal lines mark the states of constant
magnetization $M=2I-i$, where we refer to the maximal value
on such a diagonal as $m_{max}$. Obviously, we have 
$m_{max}=I-1/2$ for $i=2I$ and $m_{max}=I+1/2$ 
otherwise. Following a line of constant magnetization starting from $m_{max}$, 
one recognizes that
from a certain value $m=m_{min}$ on, all occurring states result from those 
with larger values of $m$ by 
interchanging the respective magnetizations $(m,M-m)$. 
In Fig. \ref{Fig:State} these ``complementary''
states are connected by dotted bended lines. 
It is easy to see that if $i$ is odd, we have 
$m_{min}=(2I-i)/2$, whereas for an even value of $i$ 
we have to add $(1/2)$ so that $m_{min}=(2I-i+1)/2$. It is now a simple fact 
that there are states which do not have a complement. 
This is the case for the states with $m=m_{max}$ 
if $i \neq 2I$ and for those with $m=m_{min}$ provided $i$ is odd or equal 
to zero. 

With respect to later considerations it turns out to be more convenient to 
use an ansatz which is a sum over pairs of complementary states, 
rather than a direct superposition of the 
states (\ref{symmst}). Hence we introduce coefficients 
$\alpha_m, \alpha'_m$ for any state with
$m\geq m_{min}$ and its complement and combine them to a 
sum running from $m_{min}$ to $m_{max}$:
\begin{eqnarray}
\label{ansatz}
\nonumber \fl \ket{\pm,i} := \sum_{m=m_{min}}^{m_{max}} 
\left[ \alpha_m \left( \ket{+,m}\ket{-,2I-i-m} 
\pm \ket{-,2I-i-m}\ket{+,m} \right) \right. \\
+\left. \theta(m) \alpha'_m \left( \ket{+,2I-i-m}
\ket{-,m} \pm \ket{-,m}\ket{+,2I-i-m} \right) \right]
\end{eqnarray}
The complements of the respective $m=m_{max}$ states automatically vanish, 
whereas the function $\theta(m)$ accounts for the $m_{min}$ states 
without a complement:
\begin{equation}
\theta(m)=\Theta( m  - m_{min} -1)+ \delta_{i\;\mbox{mod}\;2, 0}\delta_{m,m_{min}}
\end{equation}
where the Heavyside function $\Theta(x)$ is unity for any $x\geq 0$
and zero otherwise. 

Clearly, the ansatz (\ref{ansatz}) is an eigenstate of $H_G$,
\begin{equation}
H_G\ket{\pm,i}=-\frac{A}{2}\ket{\pm,i},
\end{equation}
consisting of triplet and singlet terms. Eliminating the latter by demanding
\numparts
\begin{eqnarray}
\label{QQ1}
\nonumber \fl \sum _{m=m_{min}}^{m_{max}} 
\biggl[  \left(\alpha_m \mu^+(m)\mu^+(2I-i-m) \mp \theta(m) 
\alpha'_m 
\mu^-(m)\mu^-(2I-i-m)\right) \biggr. \\
\left. \times \left(\ket{m-\frac{1}{2},2I-i-m+\frac{1}{2}} 
\mp \ket{2I-i-m+\frac{1}{2},m-\frac{1}{2}} \right) \right. \\ 
\nonumber + \left.\left(\alpha_m \mu^-(m)\mu^-(2I-i-m) \mp \theta(m) 
\alpha'_m \mu^+(m)\mu^+(2I-i-m)\right) \right. \\
\label{QQ2}
\left.  \times \left(\ket{m+\frac{1}{2},2I-i-m-\frac{1}{2}} 
\mp \ket{2I-i-m-\frac{1}{2},m+\frac{1}{2}} \right) \right]=0,
\end{eqnarray}
\endnumparts
we arrive at an eigenstate of $H_c$.

Let us first consider the two particularly 
simple cases $i=0$ and $i=2I$. For $i=0$, i.e. $M=2I$,  
the sum consists of only one term 
$m=I+1/2$. In this case the contributions related $\alpha'_{I+1/2}$ 
to in (\ref{QQ1}) 
and (\ref{QQ2}) are vanishing, implying that for positive parity 
the unwanted singlet
terms are automatically zero. This means that the largest degenerate 
multiplet with $J=2I$ always has positive parity, 
as demonstrated by our numerics. 

In the other case $i=2I$, i.e. $M=0$, one easily sees that
\begin{equation}
 \mu^+(m)\mu^+(-m) = \mu^-(m)\mu^-(-m)\,.
\end{equation}
If $i$ is even, this condition 
means that for every $m$ the singlet terms can be 
eliminated by simply choosing $\alpha_{m}= \pm \alpha'_m$. Therefore, 
in this case we always have an equal number of multiplets with positive and 
with negative parity. As mentioned above, for an odd value of $i$ the 
summand with $m=m_{min}=0$ does not have a complement. 
However, for positive parity the unwanted terms vanish automatically 
so that the number of positive multiplets is larger by one than the 
number of negative multiplets. In total, 
we get $2I$ solutions as suggested by our numerics. 

These solutions result from demanding that the terms in (\ref{QQ1})
and (\ref{QQ1}) vanish separately, while, strictly speaking, only their
sum is required to be zero. However, it is indeed 
simple to see that there are no further solutions: Demanding that the
total sum vanishes leads to the conditions
\begin{eqnarray*}
&&\mu^+(m)\mu^+(-m)\left(\alpha_m \mp \alpha'_m \right) \\
&&=\pm \mu^+(m-1)\mu^+(-m+1)\left(\alpha_{m-1} \mp \alpha'_{m-1}\right),
\end{eqnarray*}
and $\left(\alpha_{\frac{1}{2}} \mp \alpha'_{\frac{1}{2}}\right)
=\left(\alpha_{I-\frac{1}{2}} \mp \alpha'_{I-\frac{1}{2}}\right)=0$, which obviously 
give the same solutions as above. 
In summary, the resulting eigenstates at $i=2I$ ($M=0$) can be
formulated most compactly as
\begin{eqnarray}
\label{zeroM}
\nonumber \ket{\pm,2I,m}&:=&\ket{\uparrow \uparrow}
\left(\ket{m-\frac{1}{2},-m-\frac{1}{2}} 
\pm \ket{-m-\frac{1}{2},m-\frac{1}{2}}\right) \\
&-& \ket{\downarrow \downarrow}
\left(\ket{m+\frac{1}{2},-m+\frac{1}{2}} 
\pm \ket{-m+\frac{1}{2},m+\frac{1}{2}}\right)\,.
\end{eqnarray}
That $\Hh$ lies \textit{fully} in the kernel of (\ref{Comm}) becomes clear  
at this point: There are $2I$ degenerate multiplets with alternating 
parity, each of which gives one state with $M=0$. Above 
we constructed states which are superpositions of exactly those 
states and lie in kernel of (\ref{Comm}). They can be combined 
to give eigenstates of $\vec{J}^2$, so that $\Hh$ can be 
constructed simply by applying $J^{\pm}$. From 
$\left[\vec{J},H_G \right]=\left[\vec{J},H_c \right]=0  $ it follows
\begin{equation}
 \left[J^{\pm},\left[H_G,H_c \right] \right]=0,
\end{equation}
meaning that a state resulting from the application of $J^{\pm}$ to a state 
lying the kernel of (\ref{Comm}) again lies in the kernel of (\ref{Comm}). 
Therefore the full degenerate subspace is located there. 

\subsection{Complete construction}

Now we come to the construction of the full degenerate space $\Hh$. 
In an immediate approach we follow the route described above and 
combine the states (\ref{zeroM}) to eigenstates of $\vec{J}^2$ such 
that $\Hh$ can be generated by applying $J^{\pm}$. Unfortunately the 
construction of $\vec{J}^2$ eigenstates is possible only up the 
solution of a homogeneous set of equations with a (symmetric) 
tridiagonal coefficient matrix, which has to be carried out numerically. 
Due to the simple shape of the matrix such a problem has the very 
low complexity of $\mathcal{O}(2I)$ so that even systems of realistic 
size with respect to experimental situations in for example semiconductor 
quantum dots $I\sim 10^6$ can be treated on conventional computers 
\cite{Petta06, Koppens06, Hanson07, Braun05}. However, in a second approach 
we construct a basis of $\Hh$ in a fully analytical fashion. The 
resulting basis states  are eigenstates of $J^z$ and $\tau$, but they are 
neither orthogonal nor do they satisfy the $\vec{J}^2$ symmetry. 
Nevertheless, for both applied as well as more mathematical future 
considerations it will be helpful to have closed analytical expressions 
at hand.

\subsubsection{First approach: Construction of eigenstates of $\vec{J}^2$ with $M=0$}

As mentioned above, our first approach consists in using the particularly 
simple solutions for $i=2I$ given in (\ref{zeroM}) 
by combining them to eigenstates of $\vec{J}^2$
such that applying the ladder operators $J^{\pm}$ generates the full space 
$\Hh$. Hence we demand
\begin{eqnarray*}
 \vec{J}^2 \sum_{m=m_{min}}^{I-\frac{1}{2}} \beta_m 
\ket{\pm,2I,m}=J(J+1) \sum_{m=m_{min}}^{I-\frac{1}{2}} \beta_m \ket{\pm,2I,m} \\
\Leftrightarrow \vec{J}^2 \sum_{m=m_{min}}^{I-\frac{1}{2}} \beta_m 
\ket{\pm,2I,m}-J(J+1) \sum_{m=m_{min}}^{I-\frac{1}{2}} \beta_m \ket{\pm,2I,m}=0.
\end{eqnarray*}
Explicitly this reads
\numparts
 \begin{eqnarray}\label{J1}
\nonumber \ket{\uparrow \uparrow} \sum_{m=m_{min}}^{I-\frac{1}{2}} \beta_m \left[ \left( 2I(I+1) -2 (m+\frac{1}{2})(m-\frac{1}{2})-J(J+1)\right) \right. \\ 
\left. \times \left(\ket{m-\frac{1}{2},-m-\frac{1}{2}}\pm\ket{-m-\frac{1}{2},m-\frac{1}{2}} \right) \right.\\ \label{J2}
+ \left. \nu^+(m-\frac{1}{2})\nu^-(-m-\frac{1}{2}) \left(\ket{m+\frac{1}{2},-m-\frac{3}{2}}\pm\ket{-m-\frac{3}{2},m+\frac{1}{2}} \right) \right. \\ \label{J3}
+ \left. \nu^-(m-\frac{1}{2})\nu^+(-m-\frac{1}{2}) \left(\ket{m-\frac{3}{2},-m+\frac{1}{2}}\pm\ket{-m+\frac{1}{2},m-\frac{3}{2}} \right) \right] \\
\nonumber -\ket{\downarrow \downarrow} \sum_{m=m_{min}}^{I-\frac{1}{2}} \beta_m \left[ \left( 2I(I+1) -2 (m+\frac{1}{2})(m-\frac{1}{2})-J(J+1)\right) \right.\\ 
\nonumber \left. \times \left(\ket{m+\frac{1}{2},-m+\frac{1}{2}}\pm\ket{-m+\frac{1}{2},m+\frac{1}{2}} \right) \right.\\
\nonumber + \left. \nu^+(m+\frac{1}{2})\nu^-(-m+\frac{1}{2}) \left(\ket{m+\frac{3}{2},-m-\frac{1}{2}}\pm\ket{-m-\frac{1}{2},m+\frac{3}{2}} \right) \right. \\
\nonumber + \left. \nu^-(m+\frac{1}{2})\nu^+(-m-\frac{1}{2}) \left(\ket{m-\frac{1}{2},-m+\frac{3}{2}}\pm\ket{-m+\frac{3}{2},m-\frac{1}{2}} \right) \right]=0,
 \end{eqnarray}
\endnumparts
where $\nu^{\pm}(m)=\sqrt{I(I+1)-m(m\pm1)}$ and hence
\begin{eqnarray*}
\nu^+(m-\frac{1}{2})\nu^-(-m-\frac{1}{2})=\nu^+(m+\frac{1}{2})\nu^-(-m+\frac{1}{2})\\ 
\nu^-(m-\frac{1}{2})\nu^+(-m-\frac{1}{2})=\nu^-(m+\frac{1}{2})\nu^+(-m-\frac{1}{2}),
\end{eqnarray*}
which is plausible, because the $\ket{\uparrow \uparrow}$ and $\ket{\downarrow \downarrow}$ terms must vanish separately. Note that all components with $\ket{\uparrow \downarrow},\ket{\downarrow \uparrow}$ are equal to zero. It is now simple to see that the state in (\ref{J2}) for some $m$ is identical to the one in (\ref{J2}) for $(m+1)$ and to the one in (\ref{J3}) for $(m+2)$. For an even value of $i$ eliminating these terms gives the following set of equations
\begin{eqnarray}
\label{sys1}
\nonumber \fl \beta_m \left[ \nu^+(m-\frac{1}{2}) \nu^-(-m-\frac{1}{2}) \right]+\beta_{m+1} \left[2I(I+1)-2(m+\frac{3}{2})(m+\frac{1}{2})-J(J+1) \right] \\
\nonumber + \Theta(I-\frac{3}{2}-m) \beta_{m+2} \left[ \nu^+(m+\frac{3}{2}) \nu^-(-m-\frac{5}{2})\right] =0 \\
\nonumber \beta_{\frac{1}{2}}\left[2I(I+1)-J(J+1)\pm I(I+1) \right]+\beta_{\frac{3}{2}}\left[\nu^{-}(1)\nu^+(-2) \right]=0,
\end{eqnarray}
where $m=1/2,\ldots, I-3/2$. This yields a symmetric tridiagonal matrix. However, the symmetry of the matrix is destroyed if $i$ is odd. In this case we have
\begin{eqnarray}
\label{sys2}
\nonumber \fl \beta_m (1\pm\delta_{m,0}) \left[ \nu^+(m-\frac{1}{2}) \nu^-(-m-\frac{1}{2}) \right]+\beta_{m+1} \left[2I(I+1)-2(m+\frac{3}{2})(m+\frac{1}{2})-J(J+1) \right] \\
\nonumber + \Theta(I-\frac{3}{2}-m) \beta_{m+2} \left[ \nu^+(m+\frac{3}{2}) \nu^-(-m-\frac{5}{2})\right] =0 \\
\nonumber \beta_{0}\left[2I(I+1)-J(J+1)-\frac{1}{2} \right]+\beta_{1}\left[\nu^{-}(\frac{1}{2})\nu^+(-\frac{3}{2}) \right]=0,
\end{eqnarray}
where $m=0,\ldots,I-1/2$. The two above systems now have to be solved numerically for the different values of $J$. 

\subsubsection{Second approach: Explicit elimination of singlet contributions via ansatz}
Our second approach, which in contrast to the above one will lead to closed analytical expressions for the degenerate eigenstates, consists in \textit{directly} determining the constants $\alpha_m, \alpha'_m$ for every value of $M$. As already used above, considering (\ref{QQ1}) for some $m$ and (\ref{QQ2}) for $(m-1)$, one sees that the respective states become identical up to a factor $(\mp 1)$. The idea is now to eliminate these terms systematically, so that we get a sufficient number of linearly independent eigenvectors. As indicated in Fig. \ref{Fig:State} by the solid bended lines, this can be done by simply superposing an increasing number of successive terms and choosing all other constants to be equal to zero. Of course these solutions are by no means unique. We just choose the most compact ones. For an odd value of $i$ this yields the following quite cumbersome solutions
\begin{eqnarray}
\label{sol}
\nonumber  \alpha_{I+\frac{1}{2}-\lambda}= (-1)^{\kappa-\lambda} (\mp1)^{\kappa-\lambda-1} N_{\kappa} \left[  \mu^+(I+\frac{1}{2}-\lambda) \mu^+(I-i-\frac{1}{2}+\lambda)\right.\\ 
\left. \mp \mu^-(I+\frac{1}{2}-\lambda) \mu^-(I-i-\frac{1}{2}+\lambda)\right]^{-1}  ,
\end{eqnarray}
where $\lambda=0, \ldots, (\kappa-1)$ and 
\begin{small}
\begin{equation*}
\label{odd}
\fl N_{\kappa}= 
\cases{ 
\left[ \mu^-(I+\frac{1}{2}-\kappa) \mu^-(I-i-\frac{1}{2}+\kappa) - \frac{\left( \mu^+(I+\frac{1}{2}-\kappa) \mu^+(I-i-\frac{1}{2}+\kappa)\right)^2}{\mu^-(I+\frac{1}{2}-\kappa) \mu^-(I-i-\frac{1}{2}+\kappa)}\right] \alpha_{I+\frac{1}{2}-\kappa} &  \\ 
\left[(\mu^+(I-\frac{i}{2}))^2 
\mp (\mu^-(I-\frac{i}{2}))^2  \right] 
\alpha_{I-\frac{i}{2}},        & }
\end{equation*}
\end{small}
where the first line refers to $\kappa= 1, \ldots, (i-1)/2$ and the second line applies to $\kappa=(i+1)/2$.
For even $i$ and negative parity the solution coincides with the first 
line of (\ref{odd}), where now $\kappa=1, \ldots, i/2$. Considering 
positive parity we get:
\begin{small}
\begin{equation*}
\label{even}
\fl N_{\kappa}= \cases{ 
\left[ \mu^-(I+\frac{1}{2}-\kappa ) 
\mu^-(I-i-\frac{1}{2}+\kappa) - 
\frac{\left( \mu^+(I+\frac{1}{2}-\kappa) 
\mu^+(I-i-\frac{1}{2}+\kappa)\right)^2 }{\mu^-(I+\frac{1}{2}-\kappa) 
\mu^-(I-i-\frac{1}{2}+\kappa)}\right] 
\alpha_{I+\frac{1}{2}-\kappa} & \\ 
\left[ \mu^+(I-\frac{i}{2}+\frac{1}{2})
\mu^+(I-\frac{i}{2}-\frac{1}{2}) - 
\mu^-(I-\frac{i}{2}+\frac{1}{2}) 
\mu^-(I-\frac{i}{2}-\frac{1}{2}) \right]  
\alpha_{I-\frac{i}{2}+\frac{1}{2}} & }
\end{equation*}
\end{small}
with $\kappa= 1,\ldots, (i-2)/2$ for the first line and $\kappa=i/2$ for the second one. Due to the presence of an $\alpha'_{m_{min}}$ term, in contrast to the case of an odd $i$, here we have an additional solution. This results by simply choosing all constants to be equal to zero except for $\alpha_{m_{min}}$ and $\alpha'_{m_{min}}$, which are determined by eliminating the (\ref{QQ2}) term:
\begin{equation} 
\alpha'_{I-\frac{i}{2}+\frac{1}{2}} = \mp \frac{\mu^-(I-\frac{i}{2}+\frac{1}{2}) \mu^-(I-\frac{i}{2}-\frac{1}{2})}{\mu^+(I-\frac{i}{2}+\frac{1}{2}) \mu^+(I-\frac{i}{2}-\frac{1}{2})} \alpha_{I-\frac{i}{2}+\frac{1}{2}}
\end{equation}
Note that all the remaining constants are determined by the normalization condition. Let us give a brief discussion of the above results. With respect to subspaces of fixed $i$, the degeneracies shown in Figs. \ref{Fig:Spektrum1} and \ref{Fig:Spektrum2} yield the pattern shown in Tab. \ref{Tab:T1}. Obviously for any $i$ there are $(i+1)$ states. If $i$ is odd, there is an equal number of states with positive and with negative parity, whereas for an even value of $i$ the number of states with positive parity is larger by one than the number of states with negative parity. This is perfectly reproduced by our solutions: For an odd $i$ the index $\kappa$ in (\ref{odd}) runs up to $(i+1)/2$ for each parity, meaning that there are $(i+1)$ solutions in total. If $i$ is even, (\ref{even}) yields $i/2$ solutions for each parity and an additional one for positive parity.
\begin{table}[h!]
\begin{tabular}{c|l}
i & $\tau$ \\ \hline
0 & + \\
1 & +- \\
2 & +-+ \\
3 & +-+- \\
\multicolumn{2}{c}{$\ldots$}
\end{tabular}
\caption{\label{Tab:T1} Numerically detected degeneracy pattern.}
\end{table}

\section{The role of the inversion symmetry}
\label{inv}

In the preceding section we have constructed the full degenerate subspace 
by determining the coefficients in (\ref{ansatz}) so that we arrive at 
triplet states of the two central spins. Obviously, such a construction is 
still possible if inversion symmetry is broken.
Note that if $I_1 \neq I_2$, 
additional labels for the spin length have to be introduced in (\ref{Mu}) 
and (9). However, it is simple to see that in general our states are no 
longer eigenstates with respect to $H_G$, because the degeneracy between 
the $H_G$ eigenstates $\ket{\pm,m_1}\ket{\mp,m_2}$ is lifted. Indeed this can 
be easily recovered by demanding 
$E_+(A_1,I_1)+E_-(A_2,I_2)=E_-(A_1,I_1)+E_+(A_2,I_2)$, yielding the quite remarkable relation
\begin{equation}
\label{origin}
 A_1 \mathrm{dim}\left( \mathcal{H}_1\right)= A_2 \mathrm{dim}\left( \mathcal{H}_2 \right),
\end{equation}
where $\mathrm{dim} \left(\mathcal{H}_i\right)=2I_i+1$ denotes the dimension of the Hilbert space associated with $I_i$. Note that the bath spins $\sum_{j=1}^{n_i}\vec{I}_i^j$ couple to different values of $I_i$ so that, by a corresponding choice of $A_1, A_2$, several different degenerate subspaces can be constructed. 

\begin{figure}
\includegraphics[width=8cm]{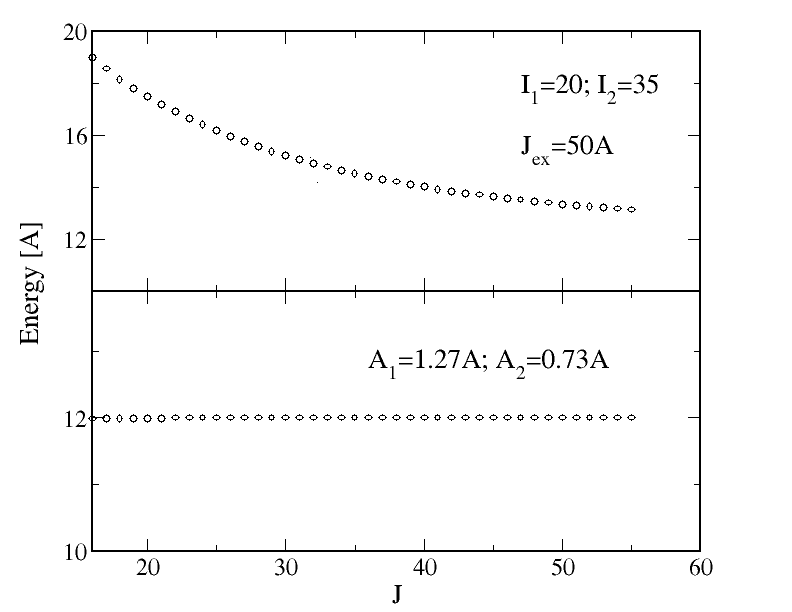}
\caption{\label{Fig:recover} Formerly degenerate branch of the spectrum for $I=20$ and $J=25A$. The inversion symmetry is broken by choosing $I_1 \neq I_2$. The deviation from the degenerate case is stronger for multiplets with a small value of the quantum number $J$ than for those with a large value. The bottom panel shows the spectrum with the degeneracy recovered by choosing $A_1 \neq A_2$ due to (22).  }
\end{figure}

Relation (\ref{origin}) means that the inversion symmetric case is only an example of a whole class of systems exhibiting the same type of systematic degeneracy. In Fig. \ref{Fig:recover} we plot the relevant part of the spectrum for $I_1 \neq I_2$ with $I_2>I_1$. In the upper panel the couplings violate (\ref{origin}) and consequently the degeneracy between the multiplets is lifted. In the bottom panel it is recovered by choosing $A_1,A_2$ according to (\ref{origin}), leading to 
\numparts
\begin{eqnarray}
A_1=\frac{1+2I_2}{I_1+I_2+1}A\\
A_2=\frac{1+2I_1}{I_1+I_2+1}A.
\end{eqnarray}
\endnumparts
In direct analogy to the inversion symmetric case the branch begins at $(I_1+I_2)$ and ends at $(I_2-I_1+1)$.

The relation (\ref{origin}) has a concrete physical meaning: Consider a semiconductor double quantum dot. Here the electron spin interacts with the surrounding nuclear spins via the hyperfine interaction, yielding a system of two coupled Gaudin models. The role of the couplings $A_1, A_2$ is played by the overall coupling strengths of the respective dots, given the sum of all hyperfine coupling constants (which depend on the properties of the respective material). The size of the Hilbert spaces results from the spatial extent of the respective electron wave function. If it is e.g. stretched over a larger area, each individual coupling decreases, but the sum remains unaltered. Hence, in an approximative sense, the relation (\ref{origin}) can always be realized by properly adjusting the electron wave function.

With respect to possible future applications of $\Hh$ it is important to note that for parameters only weakly violating (\ref{origin}), the multiplets are still \textit{nearly} degenerate: Let us fix $A_1= A_2$ and vary $I_1,I_2$ so that (\ref{origin}) is violated. From the eigenvalues of $H_G$ it is clear that the degeneracy is lifted in a continuous way. Furthermore, as can be seen very well in the upper panel of Fig. \ref{Fig:recover}, the influence of $I_1 \neq I_2$ on multiplets with small quantum numbers $J$ is much stronger than on those with large values of $J$. This is also the case if we analogously choose $A_1 \neq A_2$.

\section{Conclusion}
In summary we have reported an unexpected systematic degeneracy in an inversion symmetric system of two coupled Gaudin models with homogeneous couplings. This leads to a degenerate subspace of macroscopic size. We have constructed the complete degenerate subspace, which is fully located in the kernel of the commutator between the two Gaudin models and their coupling term. Furthermore we have studied the role of the inversion symmetry. Indeed it turns out that the inversion symmetric case is only an example for a whole family of systems all of which share the same type of systematic degeneracy. This exclusively originates in the degeneracy of two eigenspaces of the Gaudin part of the Hamiltonian, yielding a remarkable relation between the dimension of the bath Hilbert spaces and the couplings.

Nevertheless, so far we have not been able to detect the (possibly continuous)
symmetry underlying  
this remarkable degeneracy. i.e. a set of generating operators that would
connect the highly degenerate multiplets.
This question remains as an important but probably rather intricate problem for 
further studies. Furthermore it would be fruitful to study applications 
of the degenerate space especially in the context of solid state quantum 
information processing.

\ack
This work was supported by Deutsche Forschungsgemeinschaft via SFB631.

\end{document}